\documentclass[reprint,superscriptaddress,amsmath,amssymb,aps]{revtex4-1}
\usepackage{graphicx}
\usepackage{graphicx,epstopdf,color}
\usepackage{amsfonts}
\usepackage{amsmath,amssymb,mathrsfs}
\usepackage{bm}
\usepackage{pst-grad}
\usepackage[english]{babel}
\usepackage{natbib}

\begin{document}

\title{Weyl Superconductivity in UTe$_2$}

\author{Ian M. Hayes}
\thanks{These authors contributed equally to this work}
\affiliation{Department of Physics, Maryland Quantum Materials Center, University of Maryland, College Park, MD 20742, USA.}

\author{Di S. Wei}
\thanks{These authors contributed equally to this work}
\affiliation{Geballe Laboratory for Advanced Materials, Stanford University, Stanford, CA 94305, USA.}
\affiliation{Department of Applied Physics, Stanford University, Stanford, CA 94305, USA.}

\author{Tristin Metz}
\affiliation{Department of Physics, Maryland Quantum Materials Center, University of Maryland, College Park, MD 20742, USA.}

\author{Jian Zhang}
\affiliation{State Key Laboratory of Surface Physics, Department of Physics, Fudan University, Shanghai 200433, China.}

\author{Yun Suk Eo}
\affiliation{Department of Physics, Maryland Quantum Materials Center, University of Maryland, College Park, MD 20742, USA.}

\author{Sheng Ran}
\affiliation{Department of Physics, Maryland Quantum Materials Center, University of Maryland, College Park, MD 20742, USA.}
\affiliation{NIST Center for Neutron Research, National Institute of Standards and Technology, Gaithersburg, MD 20899, USA.}

\author{Shanta R. Saha}
\affiliation{Department of Physics, Maryland Quantum Materials Center, University of Maryland, College Park, MD 20742, USA.}
\affiliation{NIST Center for Neutron Research, National Institute of Standards and Technology, Gaithersburg, MD 20899, USA.}

\author{John Collini}
\affiliation{Department of Physics, Maryland Quantum Materials Center, University of Maryland, College Park, MD 20742, USA.}

\author{Nicholas P. Butch}
\affiliation{Department of Physics, Maryland Quantum Materials Center, University of Maryland, College Park, MD 20742, USA.}
\affiliation{NIST Center for Neutron Research, National Institute of Standards and Technology, Gaithersburg, MD 20899, USA.}

\author{Daniel F. Agterberg}
\affiliation{Department of Physics, University of Wisconsin-Milwaukee, Milwaukee, Wisconsin 53201, USA.}

\author{Aharon Kapitulnik}
\email{aharonk@stanford.edu}
\affiliation{Geballe Laboratory for Advanced Materials, Stanford University, Stanford, CA 94305, USA.}
\affiliation{Department of Applied Physics, Stanford University, Stanford, CA 94305, USA.}
\affiliation{Department of Physics, Stanford University, Stanford, CA 94305, USA.}
\affiliation{Stanford Institute for Materials and Energy Sciences (SIMES), SLAC National Accelerator Laboratory, 2575 Sand Hill Road, Menlo Park, CA 94025, USA.}

\author{Johnpierre Paglione}
\email{paglione@umd.edu}
\affiliation{Department of Physics, Maryland Quantum Materials Center, University of Maryland, College Park, MD 20742, USA.}
\affiliation{NIST Center for Neutron Research, National Institute of Standards and Technology, Gaithersburg, MD 20899, USA.}
\affiliation{The Canadian Institute for Advanced Research, Toronto, Ontario, Canada.}


\begin{abstract}
The search for a material platform for topological quantum computation has recently focused on unconventional superconductors. Such material systems, where the superconducting order parameter breaks a symmetry of the crystal point group, are capable of hosting novel phenomena, including emergent Majorana quasiparticles. Unique among unconventional superconductors is the recently discovered UTe$_2$, where spin-triplet superconductivity emerges from a paramagnetic normal state \cite{ran_nearly_2019}. Although UTe$_2$ could be considered a relative of a family of known ferromagnetic superconductors \cite{huxley_ferromagnetic_2015, white_unconventional_2015}, the unique crystal structure of this material and experimentally suggested zero Curie temperature pose a great challenge to determining the symmetries, magnetism, and topology underlying the superconducting state. These emergent properties will determine the utility of UTe$_2$ for future spintronics and quantum information applications. Here, we report observations of a non-zero polar Kerr effect and of two transitions in the specific heat upon entering the superconducting state, which together show that the superconductivity in UTe$_2$ is characterized by an order parameter with two components that breaks time reversal symmetry. These data allow us to place firm constraints on the symmetries of the order parameter, which strongly suggest that UTe$_2$ is a Weyl superconductor that hosts chiral Fermi arc surface states.
\end{abstract}

\maketitle



Unconventional superconductors are capable of hosting a rich variety of phenomena, but attaining any particular desirable property, like topologically protected edge states, depends on breaking the right set of symmetries at $T_c$. The superconducting state of UTe$_2$ has attracted immense attention because many observations, including a temperature-independent NMR Knight shift \cite{ran_nearly_2019}, an anomalously large upper critical field ($H_{c2}$) \cite{ran_nearly_2019, aoki_unconventional_2019}, re-entrant superconductivity \cite{ran_extreme_2019} at high fields, chiral behavior imaged by STM \cite{jiao_microscopic_2019} and a point-node gap structure \cite{metz_point-node_2019}, all point to an odd-parity, spin-triplet state. However, the key question of whether time reversal symmetry is broken remains open. A prior attempt to measure time reversal symmetry breaking (TRSB) in UTe$_2$ using muon spin relaxation was unsuccessful due to the presence of dynamic local magnetic fields \cite{sundar_coexistance_2019}. Furthermore, TRSB in UTe$_2$ would seem to be disfavored by the fact that the irreducible point group ($D_{2h}$) representations of the orthorhombic crystal symmetry of UTe$_2$ are all one component \cite{hutanu_crystal_2019}. For time-reversal symmetry to be broken in this case, symmetry requires that two separate superconducting transitions exist and there are only three other systems, UPt$_3$, Th-doped UBe$_{13}$ and PrOs$_4$Sb$_{12}$, that show multiple superconducting transitions \cite{joynt_superconducting_2002, ott_phase_1985, maple_heavy_2002}. In this study, we resolve this issue by proposing a multi-component order parameter that is experimentally supported by measurements of TRSB in UTe$_2$, as well as two distinct phase transitions in specific heat measurements. Together, these experiments allow us to strongly constrain the symmetry classification of the order parameters to two unique candidates. 

\begin{figure*}
\includegraphics[width=15cm]{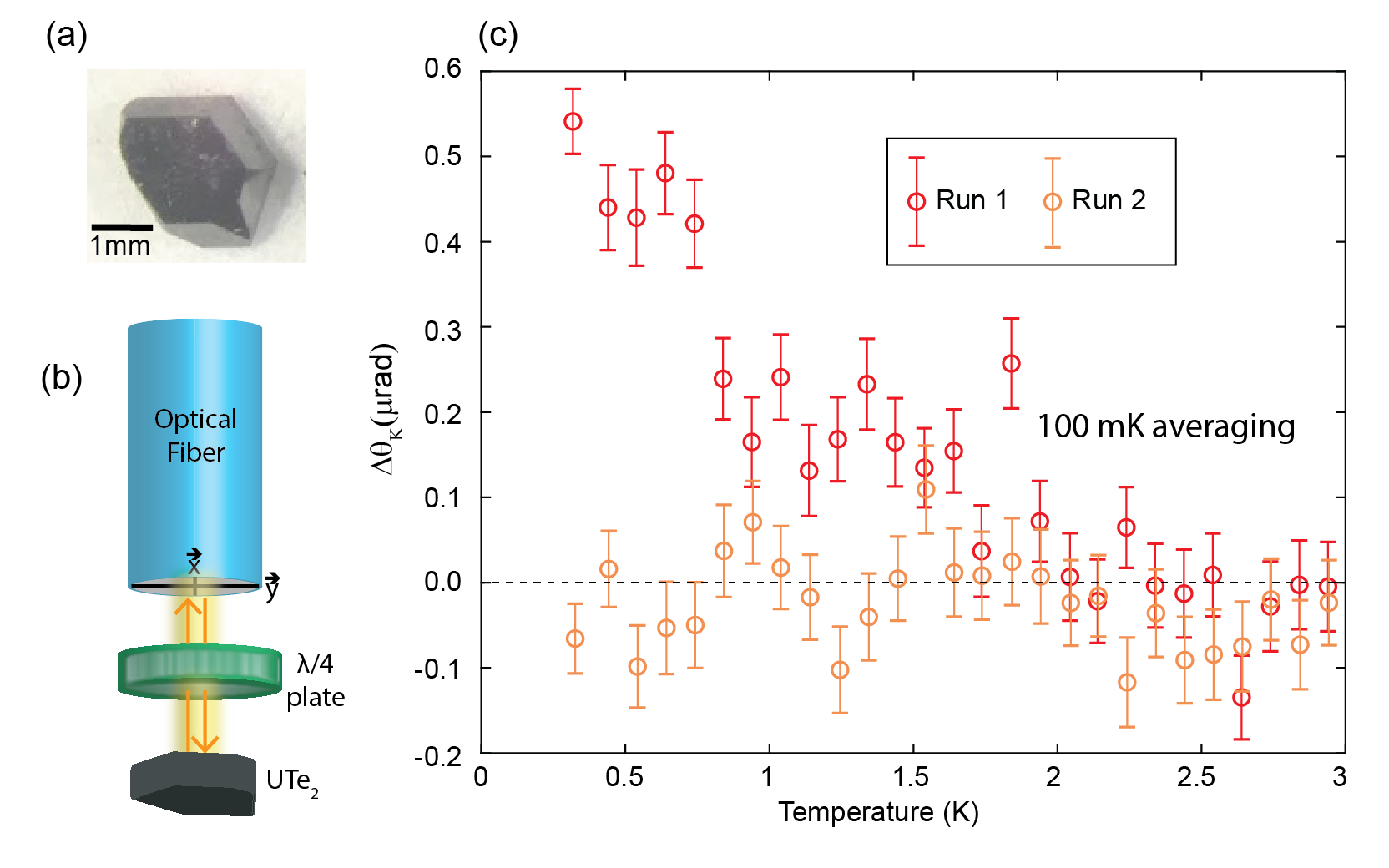} 
\caption{{\bf The evolution of the Polar Kerr angle with temperature across the superconducting transition temperature in UTe$_2$.} (a) Optical image of the UT$e_2$ single crystal used in this work. (b) Schematic of the Sagnac interferometer used to measure the polar Kerr angle. The two orthogonal axes of the fiber compose the arms of the interferometer. Light from one axis is converted to circularly polarized light at the quarter-wave plate, reflected off the sample, and then converted back to linearly polarized light at the quarter-wave plate, and then transmitted into the axis orthogonal to the one from which is originated (focusing lenses are omitted for clarity). The light is reflected off of the {\it a-b} plane. (c) Kerr angle plotted as a function of increasing temperature, after the UTe$_2$ single crystal was cooled through the superconducting transition temperature (1.6K) with zero applied magnetic field. Error bars represent statistical error of hundreds of data points averaged together over 100 mK range bins (see Supplementary Material). Two separate runs are shown. Run 1 shows no change in Kerr angle as the sample is cooled through $T_c$. Run 2 shows an increase in the Kerr angle around $T_c$, saturating at ~500 nrad. }
\label{fig:Kerr1} 
\end{figure*}

To test for possible time reversal symmetry breaking (TRSB) in the superconducting state of UTe$_2$, we performed high resolution polar Kerr effects (PKE) using a Zero-Area Sagnac Interferometer \cite{xia_modified_2006} (ZASI) that probes the sample at a wavelength of 1550 nm.  In general, the Kerr effect is defined through an asymmetry of reflection amplitudes of circularly polarized light from a given material, yielding a Kerr rotation angle $\theta_K$, and is observed only if reciprocity is broken. The Kerr effect is additionally not sensitive to Meissner effects, which normally prevent the measurement of global magnetic effects, and is therefore an optimal probe of TRSB in a superconducting system.  At the same time, probing the system at frequencies ($\omega$) much larger than the superconducting gap energy ($\Delta$) will reduce a typical ferromagnetic-like signal of order $\sim$1 rad, by a factor of $(\Delta / \hbar \omega)^2 \sim10^{-7}$, yielding a typical theoretically predicted signal of about 0.1-1 $\mu$rad \cite{goryo_impurity-induced_2008, lutchyn_frequency_2009, taylor_intrinsic_2012, brydon_loop_2019, konig_kerr_2017, wysokinski_intrinsic_2013, gradhand_kerr_2013}. However, owing to the high degree of common-mode rejection of the ZASI for any reciprocal effects (e.g. linear birefringence, optical activity, etc.), we are able to detect these small signals. 

The design and operation of our interferometer is detailed in references \cite{xia_modified_2006, kapitulnik_polar_2009}, and the basic operation is as follows: polar Kerr measurements are performed with 1550 nm wavelength light (20 $\mu$W incident power) that is polarized and then directed into a two-axes polarization maintaining optical fiber that threads down into a He-3 cryostat until it reaches our UTe$_2$ sample, which is mounted to a copper stage thermally anchored to the cold finger of the cryostat. There, a beam along each axis is reflected off the UTe$_2$ crystal face (incident on the $a$-$b$ plane of the crystal) and launched back up the opposite axis of the fiber. These two beams compose the arms of the Sagnac interferometer, enclosing a zero-area loop, and are interfered with one another to produce a signal from which the Kerr angle rotation can be extracted. The UTe$_2$ single crystal used in this study and a basic schematic of the setup is shown in Fig. 1 (a-b).
Previously, this technique has been previously used to confirm TRSB in Sr$_2$RuO$_4$ \cite{xia_high_2006} with a low-temperature saturation value of the Kerr effect of $\sim$0.1 $\mu$rad. The heavy fermion Uranium-based superconductors UPt$_3$ \cite{schemm_observation_2014} and URu$_2$Si$_2$ \cite{schemm_evidence_2015}, and filled-skutterdite PrOs$_4$Sb$_{12}$ \cite{levenson-falk_polar_2018} gave a larger signal of $\sim$0.4 to 0.7 $\mu$rad, which is expected due to their strong spin-orbit interaction. Crucially, testing the apparatus with reciprocal reflecting media such as simple BCS superconductors, gold mirrors,  and the spin-singlet d-wave heavy-fermion compound CeCoIn$_5$ \cite{schemm_polar_2017}, have yielded an expected null result.   

To begin, we report the results of polar Kerr measurements performed at low temperatures on a single crystal of UTe$_2$. 
The sample was first cooled below the $T_c$ of UTe$_2$ ($\sim$1.6K) in ambient magnetic field ($H_{ext} < 0.3$ Oe), and the Kerr angle was subsequently measured as the sample was warmed above $T_c$.
We find a small ($\sim$400 nrad at 300mK), field-trainable Kerr effect that onsets near $T_c \sim$1.6K, which is consistent with a TRSB superconducting order parameter; Fig. 1 (c) shows two runs performed identically in this manner. While Run 1 shows a signal emerging around $T_c$, and saturating at $\sim$500 nrad, Run 2 shows no discernible signal. This indicates that without an applied field domains are formed in the sample that can orient in opposite directions, and give a finite signal or no signal at all, with an average signal of zero and a standard deviation dependent on the domain to beam size (10 $\mu m$) (13). 
The detection of a finite positive Kerr signal indicates that a spontaneously large domain forms upon cooling the sample, due to TRSB in UTe$_2$.  

To orient all of the domains in one direction, the sample was cooled through $T_c$ in a small applied field of +25 Gauss. Experimentally, the magnitude of this training field has been found to be on the order of $H_{c1}$ \cite{xia_high_2006}. Once the sample reaches base temperature ($\sim$300mK), the external field is removed and the Kerr angle is measured as the sample is warmed slowly up past $T_c$. Figure 2 shows a positive finite Kerr value develop around $T_c$ in this zero-field measurement. The sign of $\theta_K$ is reversed with a negative training field (-25 Gauss), indicating that the broken time-reversal symmetry shares the same symmetry as a magnetic moment. This is due to the fact that trainability with field implies a linear coupling between the field and broken time-reversal symmetric order parameter. One point of interest is that the development of the Kerr signal with temperature does not match well to the form $\theta_K \sim (1-(T/T_c)^2)$ (Fig. 2), as seen in previous Kerr measurements. This is perhaps yet another signature of the novel microscopic roots of superconductivity in UTe$_2$. 

\begin{figure}
\includegraphics[width=8.75cm]{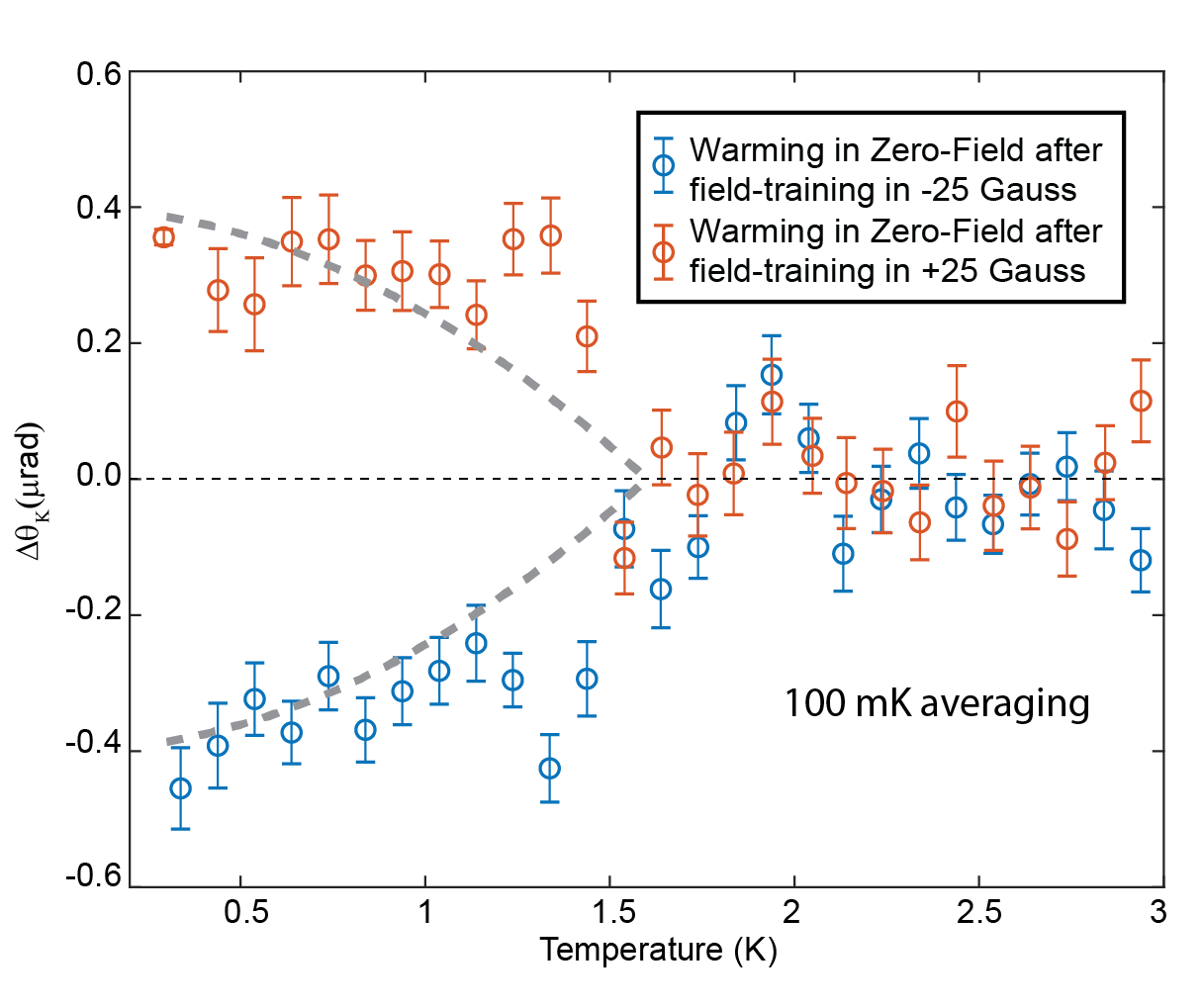} 
\caption{{\bf Magnetic field training of the Kerr effect.} Kerr angle for two different runs where the sample is warmed up past $T_c$ after being cooled in an applied field. For a positive (negative) applied field of +25 Gauss (-25 Gauss) a positive (negative) Kerr signal emerges at $T_c$ and saturates around ~400 nrad. Solid lines are guides to the eye of the form $\theta_K \sim (1-(T/T_c)^2)$.}
\label{fig:Kerr2} 
\end{figure}

As discussed above, the point group of UTe$_2$ is $D_{2h}$, which has no two-component representations. Therefore, a TRSB order parameter must be built out of two one-component representations. It is exceptionally rare for a system to support two superconducting order parameters, which might caution against our interpretation. However, we find direct evidence for the existence of two superconducting order parameters in the specific heat of UTe$_2$. Normally, superconducting states are identified by resistivity or magnetization measurements, but neighboring superconducting states would both show zero resistance and diamagnetism. For this reason, specific heat measurements have played a central role in identifying all previous examples of superconductors with multicomponent order parameters \cite{joynt_superconducting_2002, ott_phase_1985, maple_heavy_2002}. 

The specific heat in at zero field was first measured using the small-pulse method with $\Delta T= 0.5 - 2\%$ \cite{bachmann_heat_1972}. 
Four single crystals were measured from two growth batches (see the supplemental material for details). Figure 3 shows $C_p/T$ near the superconducting transition for these four samples. In each case there is a shoulder-like feature at a temperature about 75 - 100mK above the peak in $C_p/T$. This feature is quite sharp and divides the jump in the specific heat into two local maxima in the derivative, $d(C_p/T)/dT$, representing two thermodynamic anomalies. The two transitions seem to be stable to whatever perturbations are responsible for the notable difference in $T_c$ between S4 and S1, S2, and S3.  The consistent splitting of $T_c$ across four samples despite differences in growth conditions and absolute $T_c$ provides firm support for our inference that this splitting is intrinsic to UTe$_2$ and is not an artifact of inclusions or intergrowths in these crystals. Furthermore, recent work has shown that there are two well-separated transitions under pressure \cite{braithwaite_multiple_2019}. Although no splitting was seen at zero pressure, that observation still supports the idea that there are two nearly degenerate symmetry breaking possibilities for a superconducting state in UTe$_2$.

Thus, the specific heat and polar Kerr effect measurements both point to the existence of two superconducting order parameters. They also provide strong constraints on the particular irreducible representations to which they belong. Our observation that the TRSB in UTe$_2$ can be trained by a magnetic field along the crystallographic $c$-axis requires the presence of a term 
$\sim iH_c(\psi_1\psi_2^{*}-\psi_1^{*}\psi_2)$
in the free energy. Symmetry requires that this term only exists for four possibilities:\\ 

1 - $\psi_1 \in B_{3u}$ and $\psi_2 \in B_{2u}$

2 - $\psi_1 \in B_{1u}$ and $\psi_2 \in A_{u}$
 
3 - $\psi_1 \in B_{3g}$ and $\psi_2 \in B_{2g}$

4 - $\psi_1 \in B_{1g}$ and $\psi_2 \in A_{g}$\\

[See supplementary material] using the notation for irreducible representations adopted in Ref. \cite{metz_point-node_2019}. However, since UTe$_2$ is known to be a spin-triplet superconductor, we can narrow the possibilities to the first two: $B_{3u}$ and $B_{2u}$, or $B_{1u}$ and $A_u$.  

This picture can be checked by studying the two superconducting transitions as a function of magnetic field. The symmetry considerations that allow the TRSB to be trained with a field applied along the $c$-axis imply that the second transition should broaden and vanish with increasing field applied along that direction. This follows from the fact that terms like the one quoted above lead to a linear coupling between the two order parameters when a $c$-axis field is present. Symmetry considerations preclude the existence of terms like these for fields along the $a$- and $b$-axes. Therefore, we expect the two transitions to remain distinct when a magnetic field is applied along those axes, but not when a field is applied along the $c$-axis.

\begin{figure}
\includegraphics[width=8.75cm]{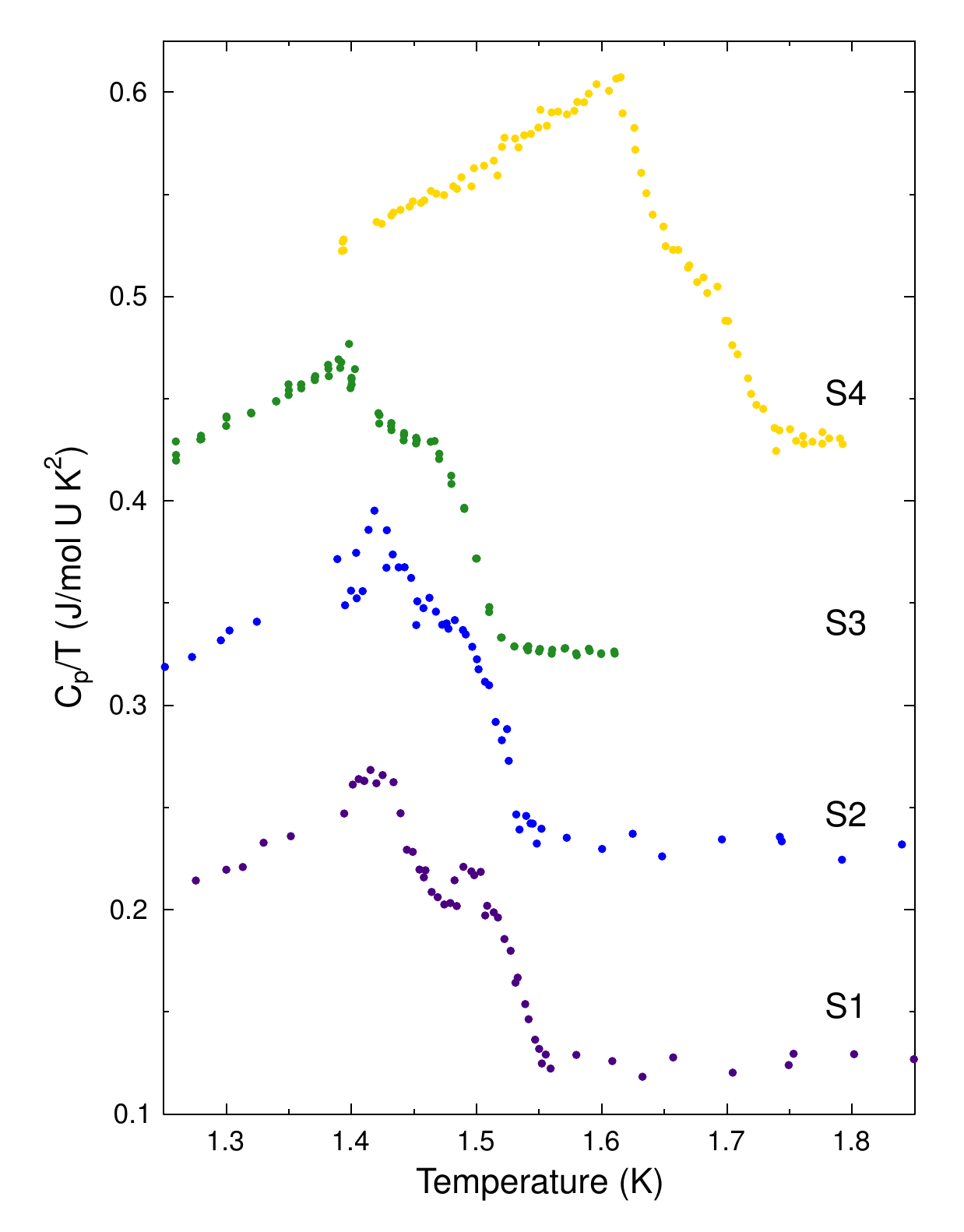} 
\caption{{\bf Superconducting transitions of UTe$_2$ in specific heat.} The specific heat over temperature per mole uranium is plotted versus temperature for four samples of UTe$_2$. The y-axis is accurate for sample S1 while the curves for the other three samples have been offset in increments of 100mJ/molK$^2$. Each sample shows two anomalies, separated by ~80mK, indicating the presence of two superconducting transitions. Samples S1-3 come from one growth batch while S4 comes from another (see SM for further details).}
\label{fig:Split_Tc1} 
\end{figure}

The field dependence of the split $T_c$ transition was measured on samples S1 and S2 by a large-pulse method (see materials and methods for details) using oriented fields up to 5T in a vector magnet. The crystal orientation was determined by measuring the field angle dependent $T_c$ for a field of 2T. For fields oriented along the $a$- and $b$-axes, the 80mK splitting remains in all fields. For field along the $c$-axis, the two transitions broaden in field so that the splitting is no longer discernible above ~1T, consistent with what we expected based on the trainability of the TRSB with a $c$-axis field. 

We thus arrive at a clear and consistent picture of a superconducting state characterized by two order parameters that belong either to $B_{3u}$ and $B_{2u}$, or $B_{1u}$ and $A_u$, and that have a relative phase of $\pi/2$, leading to a TRSB state. Next we turn to the question of the appearance of Weyl nodes.  Such nodes are topologically protected by an integer Chern number, $Z$, and have associated surface Fermi arc states \cite{sigrist_phenomenological_1991, kozii_three-dimensional_2016}. 
For an odd-parity superconducting state in a Kramer's doubly degenerate pseudo-spin band, the single-particle gaps in the quasi-particle spectrum for the two pseudo-spin-species are in general different and are given by \cite{sigrist_phenomenological_1991}
\begin{equation}
E_{\pm}=\sqrt{(\epsilon(k)-\mu)^2+|\vec{d}(k)|^2\pm |\vec{q}(k)|}
\end{equation}
where $\vec{q}(k)=\vec{d}(k)\times \vec{d}^*(k)$ denotes the non-unitary part that naturally arises when time-reversal symmetry is broken. In the two possibilities  discusses above, the gap function takes the form
\begin{equation}
\vec{d}=\vec{d}_1+i\vec{d}_2
\end{equation}
where $\vec{d}_1$ and $\vec{d}_2$ are both real. This gap function then gives rise to the effective gaps
\begin{equation}
\begin{split}
|\Delta_{\pm}|^2=|\vec{d}(k)|^2\pm |\vec{q}(k)|=|\vec{d}_1|^2+|\vec{d}_2|^2\pm 2|\vec{d_1}\times \vec{d_2}| \\=|\vec{d}_1|^2+|\vec{d}_2|^2\pm 2\sin\theta|\vec{d}_1||\vec{d}_2|
\end{split}
\end{equation}
where $\theta$ is the angle between $\vec{d}_1$ and $\vec{d}_2$. Nodes can only appear in $\Delta_-$ and for this to occur two conditions must be met:\\

\noindent i)  $\vec{d}_1\cdot\vec{d}_2=0$ ($\sin\theta=1$)\\
ii) $|\vec{d}_1|=|\vec{d}_2|$.\\
\\
In general, these two conditions will be satisfied on a line in momentum space. If this line intersects the Fermi surface, there will be a Weyl point. If it does not, the superconductor will be fully gapped. Consequently, for the two gap structures discussed above, Weyl nodes can occur at arbitrary momenta on the Fermi surface. 

\begin{figure*}
\includegraphics[width=17cm]{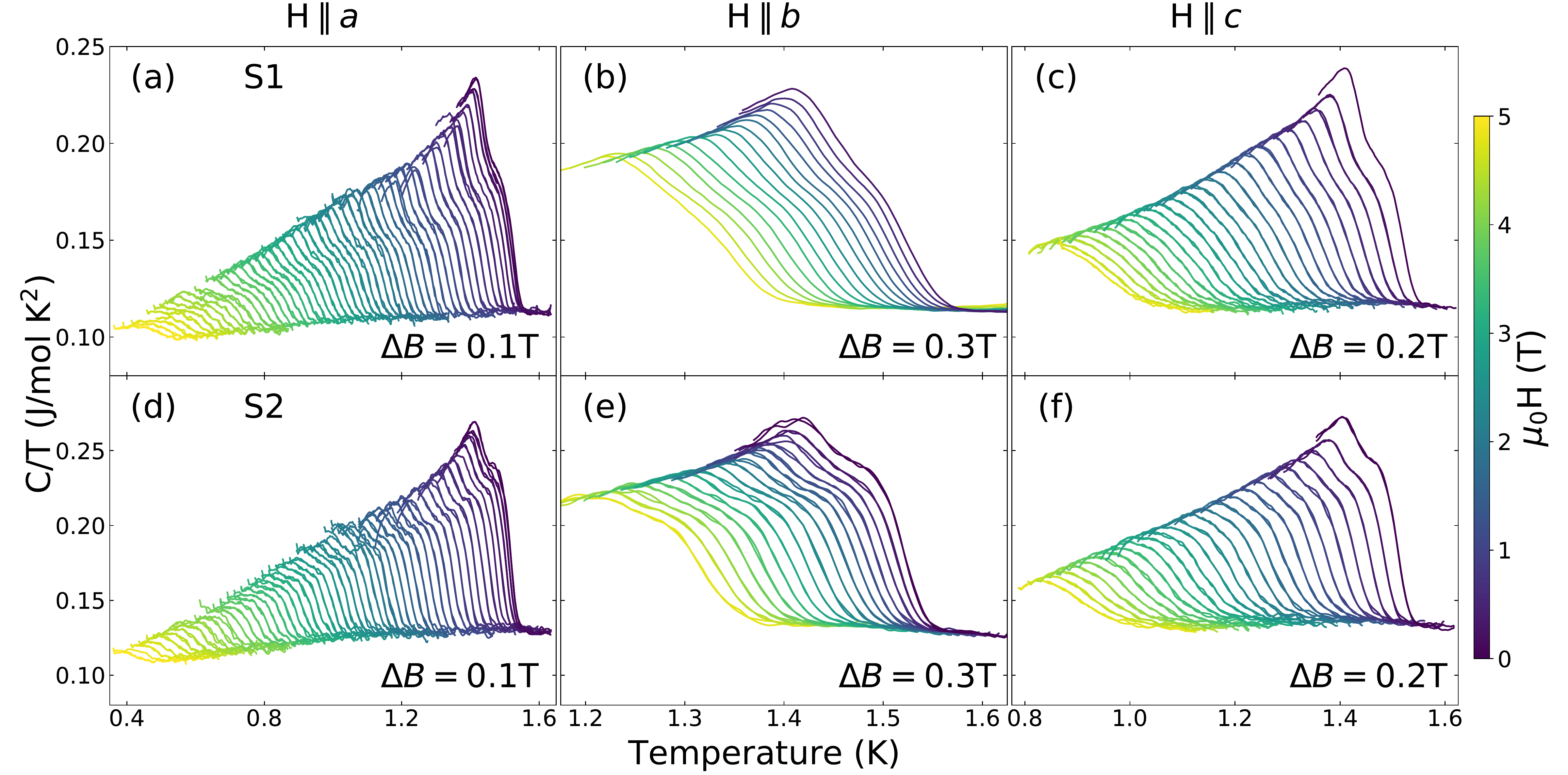} 
\caption{{\bf Magnetic field evolution of the split superconducting transition of UTe$_2$.} For samples S1 and S2, specific heat was measured by a long pulse method (see text for details) at every 100mT or 300mT along each of the crystallographic axes. Each panel corresponds to one field orientation for one of the samples and shows $C_p/T(T)$ curves gathered at each magnetic field. The curves have not been offset. In both samples, the split $T_c$ is clearly visible up to the highest measured fields when the field is oriented along the crystallographic {\it a--} or {\it b--}axes. When the field is oriented along {\it c}, however, the two transitions are indistinguishable above $\sim$2T, consistent with the linear field coupling to the product of the order parameters implied by Kerr data.}
\label{fig:Split_Tc2} 
\end{figure*}

While the above argument reveals that Weyl points can generically occur, it does not guarantee they exist. Surprisingly, it can be shown that Weyl points are expected for the $B_{2u}+iB_{3u}$ state. Given the uncertainty in the normal state  that gives rise to superconductivity, it is not possible to precisely identify the momentum dependence of the gap structure. However, symmetry places constraints on this. The symmetry dictated form of the of the corresponding gap functions are  $\vec{d}_{B2u}=f_{z,2}({\bf k})\hat{x}+f_{2,u}({\bf k})\hat{y}+f_{x,2}({\bf k})\hat{z}$ and $\vec{d}_{B3u}=f_{u,3}({\bf k})\hat{x}+f_{z,3}({\bf k})\hat{y}+f_{y,3}({\bf k})\hat{z}$, where the unknown functions $f_{x,i},f_{y,i},f_{z,i},f_{u,i}$ share the same symmetry properties as $k_x,k_y,k_z,k_xk_yk_z$. To find Weyl points for such a gap it is helpful to use insight for the Weyl semi-metal MoTe$_2$ where it was found that Weyl points appear in mirror planes \cite{wang_mote2_2016}. In particular, consider $k_x=0$, then $f_{x,i}=f_{u,i}=0$ by symmetry. This immediately implies $\vec{d}_{B2u}\cdot \vec{d}_{B3u}=0$ in this mirror plane. Furthermore, the nodal condition $|\vec{d}_{B2u}|=|\vec{d}_{B3u}|$ implies that 
$\tilde{g}_z\equiv f_{z,2}^2-f_{z,3}^2=f_{y,3}^2$. It is also possible to carry out a similar analysis for the mirror plane $k_y=0$, for which $\vec{d}_{B2u}\cdot \vec{d}_{B3u}=0$ is also satisfied. In this case, $|\vec{d}_{B2u}|=|\vec{d}_{B3u}|$ implies 
$-\tilde{g}_z2=f_{x,2}^2$. The relative minus sign in the two expressions $\tilde{g}_z=f_{y,3}^2$ ($k_x=0$) and $-\tilde{g}_z2=f_{x,2}^2$ ($k_y=0$) ensure Weyl point will exists, provided that the cross section of the Fermi surface in both the $k_x$ or $k_y=0$ planes have a circular topology (to see, this note that $\tilde{g}_z=0$ for $k_z=0$, $f_{x,i}=0$ for $k_x=0$ and $f_{y,i}=0$ for $k_y=0$ so that one of these two expressions must be satisfied somewhere on a closed Fermi surface encircling the origin in the $k_z-k_x$ or $k_z-k_y$ planes). Recent ARPES experiments suggest that such a Fermi surface exists \cite{miao_low_2019}, revealing a likely $f$-electron derived Fermi surface surrounding the $Z$ point in the Brillouin zone. Consequently, this state will give rise to at least four Weyl points in either the $k_x=0$ for the $k_y=0$ planes. A similar analysis for the $A_u + iB_{1u}$ state is given in Supplemental Materials.  

The non-unitary nature of the superconducting phase and likelihood of Weyl points make UTe$_2$ a uniquely exotic superconductor. It is highly striking that UTe$_2$ shows non-unitary superconductivity without the presence of inversion or time-reversal symmetry breaking in the normal state. This likely points to pairing mediated by ferromagnetic fluctuations, supporting the idea that UTe$_2$ is a nearly ferromagnetic system \cite{ran_nearly_2019, aoki_superconductivity_2014}. The likely Weyl points in the superconducting phase give rise to surface Fermi arc states that provide a potential explanation for the observation of chiral surface states\cite{jiao_microscopic_2019}. This study therefore opens up the possibility of topological quantum computing using UTe$_2$, as well as the discovery of a number of superconducting analogues to phenomenon in Weyl semimetals, including Fermi arcs and unusual Hall effects \cite{nagata_survey_1999}. 

\begin{acknowledgements} We thank E. Schemm, S. Tomarken, Philip Brydon, Tatsuya Shishidou, and Michael Weinert for useful discussions. Funding: Work at Stanford University was supported by the Department of Energy, Office of Basic Energy Sciences, under contract no. DE-AC02-76SF00515 and the Gordon and Betty Moore Foundation through Emergent Phenomena in Quantum Systems (EPiQS) Initiative Grant No. GBMF4529. Research at the University of Maryland was supported by the Air Force Office of Scientific Research Award No. FA9550-14-1-0332 (support of T.M.), the Department of Energy Award No. DE-SC-0019154 (specific heat experiments), the National Science Foundation Division of Materials Research Award DMR-1905891 (support of J.C.), the Gordon and Betty Moore Foundation’s EPiQS Initiative through Grant No. GBMF9071 (materials synthesis), NIST, and the Maryland Quantum Materials Center. S.R.S acknowledges support from the National Institute of Standards and Technology Cooperative Agreement 70NANB17H301. D.S.W. acknowledges support from the Karel Urbanek Fellowship in Applied Physics at Stanford University. Author contributions: D.S.W., I.M.H., A.K and J.P. conceived and designed the experiments. S.R. and N.B. synthesized the UTe2. S.R., S.R.S., and J.C. helped characterize the samples. D.S.W., J.Z. performed the polar Kerr effect measurements. I. M. H., T.M. and Y.S.E. performed the specific heat measurements. D.F.A. provided the theoretical analysis. I.M.H., D.S.W., T.M., S.R., N.B., D.F.A., J.P., and A.K. analyzed the data and wrote the paper.
\end{acknowledgements}

\bibliography{UTe2_master_bib}

\newpage
\appendix 

\section{Materials and Methods}

\subsection{Materials}

Single crystals of uranium ditelluride were grown by a chemical vapor transport method reported previously (1). Details of the characterization of these crystals were also reported in reference (1). Further details about the specific crystals selected for this study are included below.

\subsection{Methods} 

{\it Kerr Measurement data analysis}\\

Kerr angle data was acquired at a rate of 1 sample per second. Due to the slow time constant of the RF lockin amplifier used to measure the signal, each data point is correlated with its $\sim$8 closest neighbors. In order to accurately compute error bars, we eliminated these correlations by dividing the data into 50-point “chunks” and averaging the points in each chunk, leaving us with a set of averages $\{x_i\}$. These averages are then almost completely uncorrelated and thus represent independent samples with 50s of averaging time. Before plotting the data vs temperature, the chunk averages are then binned into temperature bins, with all the $x_i$ in each bin averaged together. The points plotted in all $\theta_K(T)$ graphs are these bin averages. The error bars given are the 1-$\sigma$ standard error of the mean, computed as the standard deviation of the $x_i$ in each bin divided by the square root of the number of chunks.\\

{\it Specific Heat Measurements}\\

The specific heat of UTe$_2$ was measured in an oxford dilution refrigerator (S1, S2, and S4) or in a Quantum Design PPMS using the 3He option (S3). The measurements were taken using either the conventional ``small-pulse" relaxation time method (27) or a ``large-pulse" method. In the large-pulse method the sample is heated to well above the bath temperature and then the heat capacity is extracted from the temperature versus time curve by solving the heat flow equation: 
\begin{equation}\label{base}
    C\frac{dT}{dt}=-\int_{T_b}^{T}K(T')dT'+P(t)
\end{equation}

where $C$ is the heat capacity, $T$ is the sample/platform temperature, $T_b$ is the bath temperature, $K$ is the thermal conductance between the platform and the bath, and $P$ is the power input from the heater. $dT/dt$ for the cooling curve (P=0) is determined by a second order polynomial fitting of several small temperature windows. $K$ is measured first using the small-pulse method. Because this ``large-pulse” method allowed data to be collected much faster, it was used for the dense collection of field sweeps shown in figure 4 of the main text. Figure S1 shows a comparison of the two methods, confirming that the two transitions can be seen in both measurements and  occur at the same temperatures. 

\begin{figure}
\includegraphics[width=8.75cm]{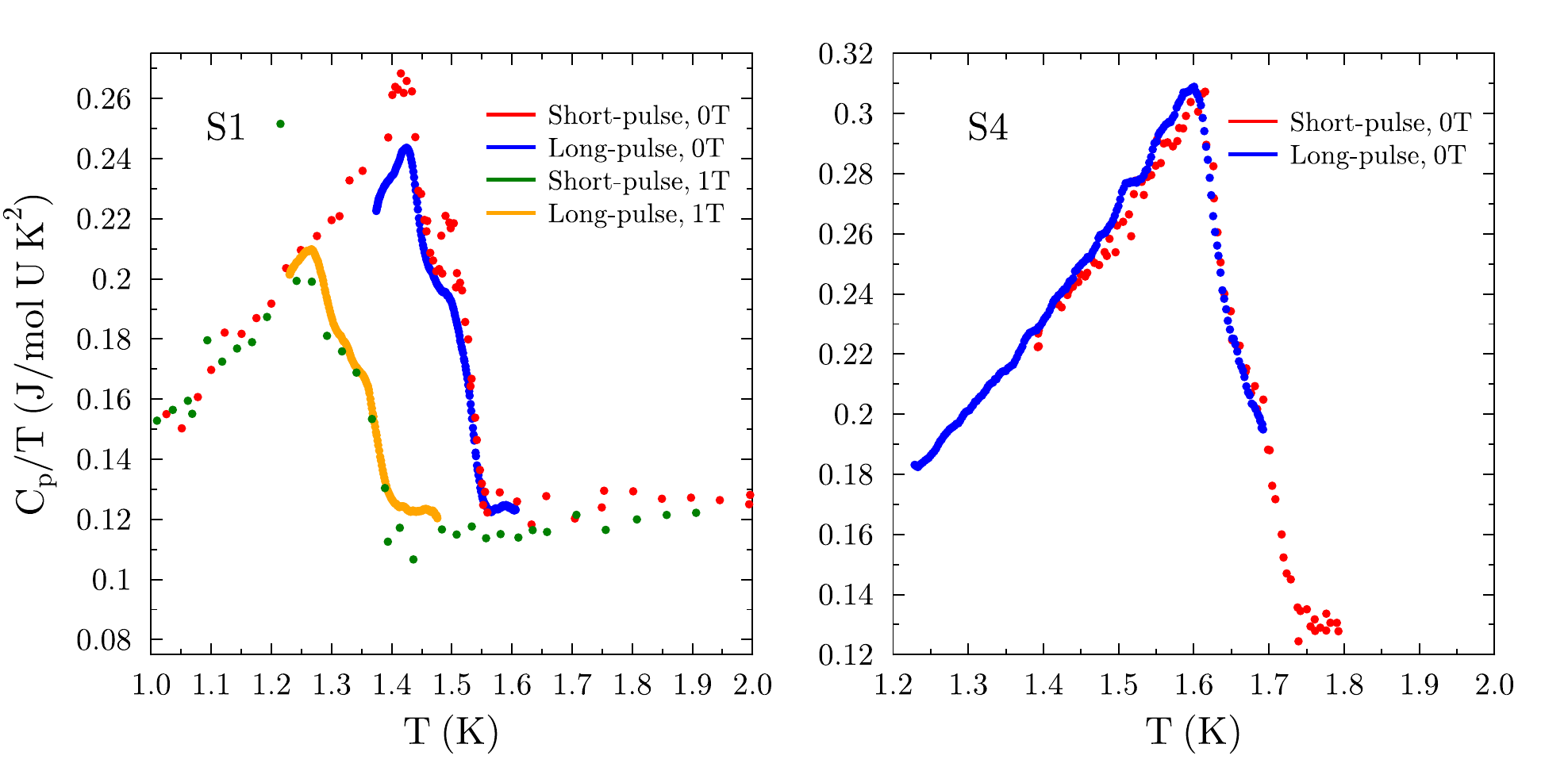} 
\caption{{\bf Short- and long-pulse specific heat measurements.} The two methods used for measuring the specific heat in this study showed good agreement. Most importantly, the two methods both show two features near the superconducting transition at the same temperatures. A few examples are shown above of this agreement are shown above.}
\label{fig:RvT} 
\end{figure}


\section{Choice of UTe$_2$ samples}

Here we present a few further details about the samples chosen for specific heat measurements in this study. Establishing the presence of an intrinsic splitting of a superconducting transition is hard because of the possibility of sample inhomogeneity. Therefore, it was important that this study use the cleanest crystals that could be made. Samples S1, S2, and S3 were taken from a single batch that showed the cleanest crystals based on the residual resistivity ratio: the ratio between the room temperature resistivity and the zero-temperature limit of the resistivity (usually taken to be the resistivity just above $T_c$ in a superconductor). Generally, larger values of this ratio indicate a smaller concentration of defects, because the defect density is the major determinate of the resistivity of a crystal in the zero temperature limit. The residual resistivity ratio (RRR) has limits as a measure of crystal quality, but in the absence of extensive x-ray characterization or electron microscopy measurements it is one of the better indicators available.

The residual resistivity ratio for the crystals from this growth was 35-40, which is as high as any RRR reported for this system (1, 4). For reasons unrelated to this study, this particular growth was done with isotopically purified Te-128. Although it is unlikely that a more uniform nuclear mass is responsible for the change in crystal quality, it is reasonable to suppose that differences in the preparation of this isotopically purified tellurium resulted in a smaller concentration of impurities relative to our usual tellurium source. Figure S2 shows a sample resistivity curve for this batch. It is worth emphasizing that the resistive transition at $T_c$ is quite sharp, making it unlikely that there are regions of the sample that have different $T_c$s. A similar point can be made about the width of the superconducting transition in the specific heat, which is quite sharp in all of these samples (see figure 3 of the main text). 

\begin{figure}
\includegraphics[width=8.75cm]{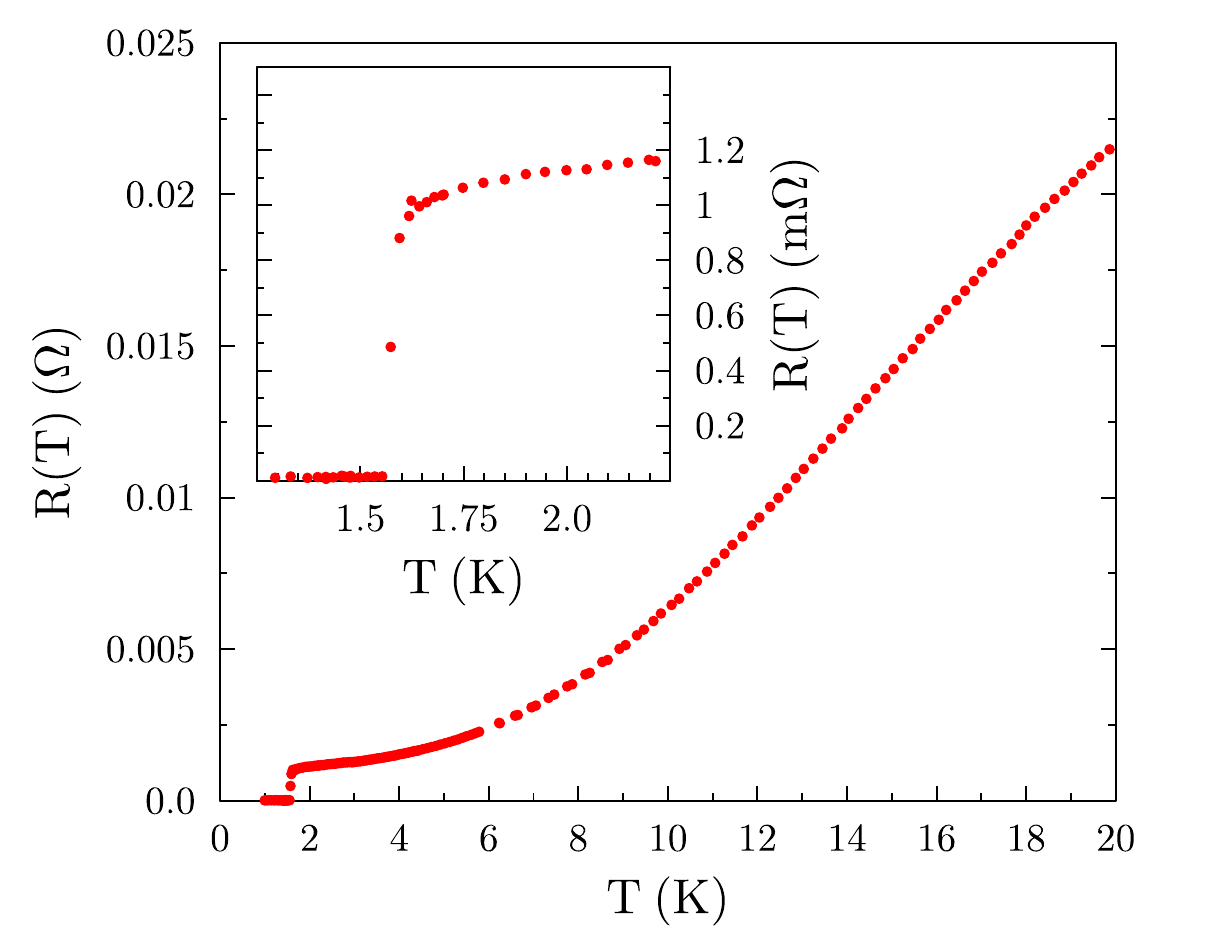} 
\caption{{\bf Resistance of UTe$_2$.} A characteristic resistance versus temperature curve for a sample from the same batch as S1, S2, and S3, confirming that the superconducting transition is sharp and at the same temperature in the resistivity and the specific heat.}
\label{fig:RvT} 
\end{figure}

Of course, it is important to establish that the splitting of Tc is not particular to one growth procedure. For that reason we measured another sample, S4. This sample comes from a typical growth done with the natural tellurium isotope abundances. Although its $T_c$ is somewhat higher than average, it is not outside of the usual range of $T_c$ that we obtain. If anything, the fact that S4 has a notably different $T_c$ strengthens that case that the two transitions are intrinsic, since they survive whatever material differences are responsible for this difference. We wish to emphasize that these four samples are the only samples that we have measured with enough point density to resolve the splitting of $T_c$ . 

\newpage
\section{Additional Theoretical Notes}

{\bf General Phenomenological Theory}\\

\vglue 0.25 cm
\begin{table*}
\begin{tabular}{|c|c|c|c|c|c|c|c|c|}
  \hline
  Irrep & $E$& $C_{2z}$& $C_{2y}$&$C_{2x}$&linear&quadratic [$\psi({\bf k})]$&$\vec{d}({\bf k})$&nodes\\
  \hline
  $A_{1g}$& 1& 1&1&1&-&$k_x^2,k_y^2,k_z^2$&-&-\\
  $B_{1g}$&1&1&-1&-1&$H_z$&$k_xk_y$&-&line\\
  $B_{2g}$&1&-1&1&-1&$H_y$&$k_xk_z$&-&line\\
  $B_{3g}$&1&-1&-1&1&$H_x$&$k_yk_z$&-&line\\
  $A_u$&1&1&1&1&-&-&$\hat{x}k_x,\hat{y}k_y,\hat{z}k_z$&-\\
  $B_{1u}$&1&1&-1&-1&$k_z$&-&$\hat{x}k_y,\hat{y}k_x,\hat{z}k_xk_yk_z$&point\\
  $B_{2u}$&1&-1&1&-1&$k_y$&-&$\hat{x}k_z,\hat{y}k_xk_yk_z,\hat{z}k_x$&point\\
  $B_{3u}$&1&-1&-1&1&$k_x$&-&$\hat{x}k_xk_yk_z,\hat{y}k_z,\hat{z}k_y$&point\\
  \hline
\end{tabular} 
\caption{Irreducible representations and representative functions for point group $D_{2h}$.}
\end{table*}

To develop a phenomenological model for the superconducting state, a central input  is the $D_{2h}$  point group symmetry of UTe$_2$  and the corresponding  irreducible  representations (REPS). These REPS are listed in Table 1 along with some representative gap functions and magnetic field orientations (denoted here as $H_i$). To relate to the structure of UTe$_2$, we take $\hat{x}$ to be along the $a$-axis,  $\hat{y}$ to be along the $b$-axis, and $\hat{z}$ to be along the $c$-axis. 

Since the REPS are all one-dimensional, any time-reversal symmetry breaking must be due to the presence of two REPS. In zero field, the Ginzburg Landau free energy density, $f$, for two REPS is generic and takes the form
\begin{equation}
\begin{split}
\tilde{f}=\alpha_1|\psi_1|^2+\alpha_2|\psi_2|^2+\frac{\beta_1}{2}|\psi_1|^4+\frac{\beta_2}{2}|\psi_2|^4+ \\ \beta_{m1}|\psi_1|^2|\psi_2|^2+\beta_{m2}(\psi_1^2\psi_2^{*2}+\psi_1^{*2}\psi_2^2).
\end{split}
\end{equation}
A key parameter in this theory is $\beta_{m2}$ since this determines whether or not time-reversal symmetry is broken below a second transition. If $\beta_{m2}>0$, then the relative phase between $\psi_1$ and $\psi_2$ will be $\pi/2$, and time-reversal symmetry is broken. If $\beta_{m2}<0$, then the relative phase will the $0$, and time-reversal symmetry is not broken. Here we take $\beta_{m2}>0$ and discuss the possible physical origin of this later. Defining $\beta_m=\beta_{m1}-2\beta_{m2}$, we can slightly simplify the free energy density
\begin{equation}
\begin{split}
f=\alpha_1|\psi_1|^2+\alpha_2|\psi_2|^2+\frac{\beta_1}{2}|\psi_1|^4+\frac{\beta_2}{2}|\psi_2|^4+ \\ \beta_{m}|\psi_1|^2|\psi_2|^2.
\end{split}
\end{equation}
There are some conditions on this free energy. In particular, if all transitions are observed to be second order then $\beta_1>0$, $\beta_2>0$ and $\beta_m^2<\beta_1\beta_2$ (this ensures that the order parameters do not diverge). If we assume that the upper phase transition has $\psi_1\ne 0$ (and $\psi_2=0$), so that this transition is given by $\alpha_1=0$ (this defines $T_{c1}$), then the second transition, for which $\psi_2
\ne 0$, is given by the condition 
\begin{equation}
\alpha_2-\frac{\beta_m}{\beta_1}\alpha_1=0
\end{equation}
this shows that the $T_{c2}$ for the second transition is shifted (this can be a shift upwards or downwards since $\beta_m$ can be positive or negative) from the original $\tilde{T}_{c2}$ (which is given by $\alpha_2=0$). While this free energy has a number of unknown coefficients which restricts its usefulness, it is possible to find one constraint that is given by experiment. In particular, the ratio of the specific jumps can be expressed as
\begin{equation}
\frac{\Delta C_2/T_{c2}}{\Delta C_1/T_{c1}}=\frac{(\beta_1-\beta_m)^2}{\beta_1\beta_2-\beta_m^2}
\end{equation}\\

{\bf Constraints from polar Kerr effect}\\

For two REPS, there is only one bilinear combination that allows for broken time-reversal symmetry. This bilinear takes the form $i(\psi_1\psi_2^*-\psi_2\psi_1^*)$. Since the Kerr measurements reveal that a $c$-axis field trains the Kerr signal, this implies the existence of a coupling term $iH_z(\psi_1\psi_2^*-\psi_2\psi_1^*)$. This coupling term can only exist if the bilinear $i(\psi_1\psi_2^*-\psi_2\psi_1^*)$ has the same symmetry as $H_z$, that is it must have a $B_{1g}$ symmetry. This restricts the order parameters $\psi_1$ and $\psi_2$ to one of four possibilities:\\

\noindent 1- $\psi_1 \in B_{3u}$ and  $\psi_2 \in B_{2u}$\\
2- $\psi_1 \in B_{1u}$ and  $\psi_2 \in A_{u}$\\
3- $\psi_1 \in B_{3g}$ and  $\psi_2 \in B_{2g}$\\
4- $\psi_1 \in B_{1g}$ and  $\psi_2 \in A_{1g}$\\

The observation of critical fields that far exceed the Pauli limiting field and the observation of a nearly constant Knight shift in UTe$_2$ suggest that only possibilities 1 or 2 are realized in UTe$_2$.\\ 

{\bf Do two transitions survive in finite field?}\\

This question can be answered by extending the Ginzburg Landau analysis to finite fields. This expansion takes the same form for all four possibilities listed above. There are many terms in this expansion, but it is only the terms that mix the two order parameter components that will turn the second phase transition into a crossover, these terms are:
\begin{equation}
\begin{split}
f_B=\epsilon i H_z(\psi_1\psi_2^*-\psi_2\psi_1^*)+\kappa [(D_y \psi_1)(D_x\psi_2)^*+\\(D_x \psi_1)(D_y \psi_2)^*]
\end{split}
\end{equation}
where $D_i =i \partial_i+e^* A_i$ and $A_i$ is the vector potential. This implies for a field applied the $\hat{a}$ or the $\hat{b}$ directions, two transitions can survive, but the second transition will be suppressed for the field applied along the $\hat{c}$ axis.\\

{\bf Stabilizing a non-unitary spin-triplet state}\\

For possibilities 1 or 2 a broken time-reversal state would imply a non-unitary order parameter. Such an order is known to be energetically expensive in weak-coupling theory, so it is reasonable to ask how it might be stabilized. One mechanism, variants of which have appeared in the literature for other materials \cite{walker_model_2002,amin_generalized_2019} and for UTe$_2$ \cite{nevidomskyy_stability_2020,yarzhemsky_time-reversal_2020},  considers low energy magnetic fluctuations to be responsible for this. In particular, consider a fluctuating moment along the $\hat{c}$ axis that is described with an order parameter $m$. The simplest quadratic free energy in terms of this order and its coupling to superconductivity is 
\begin{equation}
f_m=\alpha_m m^2+\gamma m i(\psi_1\psi_2^*-\psi_2\psi_1^*).
\end{equation}
Since $m$ is a fluctuating field, $\alpha_m>0$. It is possible to consider the partition function for this system and to integrate out the fluctuating moment to get a contribution to the free energy for the superconductivity. This contribution is
\begin{equation}
\delta f = \frac{-\gamma^2}{2\alpha_m}[i(\psi_1\psi_2^*-\psi_2\psi_1^*)]^2.
\end{equation}
This correction changes the coefficients in the original free energy given in  Eq. 1, these changes are $\tilde{\beta}_{m1}=\beta_{m1}
- \gamma^2/ \alpha_m$ and $\tilde{\beta}_{m2}=\beta_{m2}+\gamma^2/(2\alpha_m)$. That is, $\beta_{m1}$ is decreased and $\beta_{m2}$ is increased. The increase in $\beta_{m2}$ can compensate for the energy cost of having a non-unitary state and thereby  stabilize it. \\

{\bf Weyl nodes for the $A_u+iB_{1u}$ state}\\

In the manuscript, it was shown that Weyl nodes are generically expected for the $B_{2u}+iB_{3u}$ state. Here we consider a similar argument for the $A_u+iB_{1u}$ state. 

For a single pseudospin degenerate band, there are two conditions that must be satisfied at some point ${\bf k}$ on the Fermi surface to  have a Weyl point\\
\noindent i)  $\vec{d}_1\cdot\vec{d}_2=0$ ($\sin\theta=1$).\\
ii) $|\vec{d}_1|=|\vec{d}_2|$.\\

For the $A_u+iB_{1u}$ state, we have  $\vec{d}_1=\hat{x}f_{y,1} +\hat{y}f_{x,1} +\hat{z}c_u=(f_{y,1},f_{x,1},c_u)$ and $\vec{d}_2 = i(\hat{x}f_{x,2}+\hat{y}f_{y,2}+\hat{z}f_{z})=i(f_{x,2},f_{y,2}, f_{z})$ where the $\alpha_i$, the $\beta_i$, and the functions $f_i$ and $c_i$ are all real. In addition $f_{x},f_{y},f_{z,i}$ share the same transformation properties as $k_x,k_y,k_z$ and $c_{u,i}$ shares the same transformation properties as $k_xk_yk_z$. To look for Weyl points, it is useful to consider the mirror planes given by $k_x=0$ and $k_y=0$.\\

\noindent \underline{\it $k_x=0$}

For this plane, $\vec{d}_1\cdot \vec {d}_2=0$ is satisfied, therefore we only require condition ii) to be satisfied on the Fermi surface. This implies $f_{y,1}^2=f_{z}^2+f_{y,2}^2$, or $\tilde{f}_y^2=f_{y,1}^2-f_{y,2}^2=f_z^2$. This can only be satisfied if $\tilde{f}_y^2\ge 0$, if this is true everywhere on the Fermi surface, then  Weyl points will likely occur. This follows because $f_{y,i}=0$ when $k_y=0$ and $f_{z,i}=0$ when $k_z=0$, so if the Fermi surface in the $k_y$-$k_z$ plane is circular in topology and crosses the lines $k_y=0$ and $k_z=0$, there will be a Weyl point.\\

\noindent \underline{\it $k_y=0$}

For this plane, $\vec{d}_1\cdot \vec {d}_2=0$ is true by symmetry, therefore we again only require condition ii) to be satisfied on the Fermi surface. This implies $f_{x,1}^2=f_{x,2}^2+f_{z}^2$, or $\tilde{f}_x^2=f_{x,1}^2-f_{x,2}^2=f_z^2$. This can only be satisfied if $\tilde{f}_x^2\ge 0$, if this is true everywhere on the Fermi surface, then  Weyl points will occur. This follows because $f_{x,i}=0$ when $k_x=0$ and $f_{z}=0$ when $k_z=0$, so if the Fermi surface in the $k_x$-$k_z$ plane is circular in topology and crosses the lines $k_x=0$ and $k_z=0$, there will be a Weyl point.\\

Note that in this case, Weyl points can occur in either or both the $k_x=0$ and the $k_y=0$ planes. However, it is possible that no Weyl points occur for this state as well.
The Weyl points will generically carry charge $\pm 1$ and will appear in groups of four for these high symmetry planes. 

\end{document}